\documentclass[reprint, amsmath, amssymb, prl, nofootinbib]{revtex4-1}

\usepackage[utf8]{inputenc}

\usepackage[usenames, dvipsnames]{color}

\usepackage{mathrsfs}
\usepackage{graphicx}% Include figure files
\usepackage{dcolumn}% Align table columns on decimal point
\usepackage{bm}% bold math
\usepackage{slashbox}
\usepackage{color}
\usepackage{colortbl}
\usepackage[table]{xcolor}
\usepackage[normalem]{ulem}% for striking out: \sout{}

\usepackage[skip=2pt,justification=justified,singlelinecheck=off]{caption} % reduces space between figure and caption

\usepackage[colorlinks=true,
            linkcolor=blue,
            urlcolor=blue,
            citecolor=blue]{hyperref}

\usepackage[bottom]{footmisc}

\newcommand{\ie}{{i.e.~}}

\newcommand{\eg}{\textsl{e.g.~}}

\newcommand{\dd}{\mathrm{d}}
\newcommand{\sss}[1]{{\scriptscriptstyle{#1}}}

\newcommand{\uPl}{\mathrm{Pl}}

\newcommand{\uend}{\mathrm{end}}

\newcommand{\uS}{\mathrm{S}}

\newcommand{\usssS}{\sss{\uS}}

\newcommand{\usssPl}{\sss{\uPl}}

\newcommand{\nS}{n_\usssS}

\newcommand{\uNL}{\mathrm{NL}}

\newcommand{\calP}{\mathcal{P}}

\newcommand{\Mp}{M_\usssPl}

\newcommand{\fnl}{f_\uNL}

\newcommand{\rdec}{r_{\rm dec}}

\newcommand{\efolds}{$e$-folds}

\newcommand{\beq}{\begin{equation}}
\newcommand{\eeq}{\end{equation}}
\newcommand{\bea}{\begin{equation}\begin{aligned}}
\newcommand{\eea}{\end{aligned}\end{equation}}

\newlength{\wsingfig}
\newlength{\wdblefig}
\newlength{\wquadfig}
\newlength{\wtriplefig}
\newcommand{\Eq}[1]{Eq.~(\ref{#1})}
\newcommand{\Eqs}[1]{Eqs.~(\ref{#1})}
\newcommand{\Fig}[1]{Fig.~{\ref{#1}}}

\newcommand{\Ref}[1]{Ref.~{\cite{#1}}}
\newcommand{\Refs}[1]{Refs.~{\cite{#1}}}

\newcommand{\uniform}{\mathcal{U}}
\newcommand{\normal}{\mathcal{N}}

\begin{document}

% Titles proposals:
% New/Greater predictive power in the curvaton scenario
% Exploring pre-inflationary history with an stochastic spectator
% Measuring the duration of inflation (conditioned to measuring $\fnl=-5/4$)

\title{Measuring the duration of inflation with the curvaton}

\author{Jes\'{u}s Torrado$^{1}$,}
\email{jesus.torrado@sussex.ac.uk}
\author{Christian T.~Byrnes$^{1}$}
\email{c.byrnes@sussex.ac.uk}
\affiliation{${}^{1}$Department of Physics \& Astronomy, University of Sussex, Brighton BN1 9QH, United Kingdom}
\author{Robert J.~Hardwick$^{2}$}
\email{robert.hardwick@port.ac.uk}
\author{Vincent Vennin$^{2,3}$}%
\email{vincent.vennin@port.ac.uk}
\author{David Wands$^{2}$}%
\email{david.wands@port.ac.uk}
\affiliation{%
  ${}^{2}$Institute of Cosmology \& Gravitation, University of Portsmouth, Dennis Sciama Building, Burnaby Road, Portsmouth, PO1 3FX, United Kingdom
}%
\affiliation{${}^{3}$Laboratoire Astroparticule et Cosmologie, Universit\'e Denis Diderot Paris 7, 10 rue Alice Domon et L\'eonie Duquet, 75013 Paris, France}

\date{\today}

\begin{abstract}
  Simple models of single-field inflation in the very early universe can generate the observed amplitude and scale dependence of the primordial density perturbation, but models with multiple fields can provide an equally good fit to current data. We show how future observations will be able to distinguish between currently favoured models. If a curvaton field generates the primordial perturbations after inflation, we show how the total duration of inflation can be measured.
\end{abstract}

\maketitle

\section{Introduction}

In the standard cosmological model, all structures in our Universe originate from quantum vacuum fluctuations 
% stretched by the expansion and amplified by gravitational instability
\cite{Mukhanov:1981xt, Mukhanov:1982nu, Starobinsky:1982ee, Guth:1982ec, Hawking:1982cz, Bardeen:1983qw} during an early phase of accelerated expansion~\cite{Starobinsky:1980te, Sato:1980yn, Guth:1980zm, Linde:1981mu, Albrecht:1982wi, Linde:1983gd}. With the advent of high-precision measurements \cite{Adam:2015rua, Array:2015xqh} of the temperature and polarisation anisotropies of the cosmic microwave background (CMB), we can place tight constraints on this primordial epoch called inflation. At a given physical length scale, the statistical properties of cosmological fluctuations are mostly determined by the details of the classical inflationary dynamics around the time when this scale crosses the Hubble radius during inflation. The range of scales probed in the CMB then translates into a time interval during inflation of length {$\Delta N\sim 7$, measured by the number of \efolds, $N_1-N_0 \equiv \ln(a_1/a_0)$, where $a$ is the scale factor of the Universe.} If one includes measurements of the large-scale structure (LSS) of our Universe, this window is extended, but cannot exceed the last $\sim 60$ \efolds~of inflation. Besides this lower bound (whose precise value depends on the reheating expansion history~\cite{Liddle:2003as, Martin:2006rs, Martin:2010kz, Easther:2011yq, Martin:2014nya, Martin:2016oyk, Hardwick:2016whe}), the overall duration of inflation is not known.

One way to circumvent this cosmic amnesia and to learn about larger scales, hence earlier times, is through spectator fields~\cite{Hardwick:2017fjo, Hardwick:2017qcw}, whose field displacements are sensitive to a much longer phase of the inflationary epoch and which can be observationally accessible \cite{Linde:2005yw}. 

Since current CMB measurements are compatible with single-field models of inflation (if the potential is of the plateau type)~\cite{Ade:2015lrj, Martin:2013tda, Martin:2013nzq}, such extra fields {are not} required by the data. However, from a model building perspective, they are ubiquitous in many high-energy embeddings of inflation, \eg in the context of string theory~\cite{Turok:1987pg, Damour:1995pd, Kachru:2003sx, Krause:2007jk, Denef:2007pq, Baumann:2014nda}. It is of course always possible to fit the data with complex multi-field inflationary models, but the amount of fine tuning required in these models may be large, which is why models should be compared in a Bayesian framework that correctly accounts for the waste of parameter space.

The question we ask in this Letter is therefore twofold. Are there multiple-field models of inflation that are as favoured by the data as single-field plateau inflation from a Bayesian perspective? What insight can be gained on the inflationary history in these models?

We investigate these questions in one of the simplest extensions to single-field inflation where~\cite{Linde:1996gt, PhysRevD.42.313, Enqvist:2001zp, Lyth:2001nq, Moroi:2001ct, Bartolo:2002vf} a light (with respect to the Hubble scale) energetically subdominant quadratic spectator scalar field $\sigma$, called the ``curvaton'', sources primordial density perturbations together with the inflaton field $\phi$.  Denoting the effective mass of the curvaton by $m_\sigma$, the potential is of the form\footnote{{For the sake of simplicity, we neglect self-interactions in the spectator potential; in a more general setting those terms could only be neglected for sufficiently massive spectators.}}
\bea
V\left(\phi ,\sigma\right) = U\left(\phi\right) + \frac{m^2_\sigma}{2} \sigma^2 \, .
\eea
After inflation, the inflaton field decays into radiation and the energy density contained in the curvaton field, $\rho_\sigma$, may grow relative to the background energy density, until it also decays into radiation. Assuming that no isocurvature perturbations persist~\cite{Lyth:2002my, Weinberg:2004kf, Smith:2015bln}, the total adiabatic power spectrum is given by the sum of
the power spectra of the perturbations originating from each field,
\bea
\label{ps}
{\mathcal{P}}^{\mathrm{total}}_{\zeta}={\mathcal{P}}^{\phi}_{\zeta}+ {\mathcal{P}}^{\sigma}_{\zeta} \, ,
\eea
where in the case of observational interest, $\sigma_*\ll\Mp$ {(since $\sigma_*\sim\Mp$ is disfavoured by the data, see footnote \ref{footnote},}
\bea 
\label{eq:powerspecs}
\mathcal{P}^{\phi}_{\zeta} = \frac{1}{2\epsilon_*}\left(\frac{H_*}{2\pi \Mp}\right)^2
 \;\; \mathrm{and} \;\;
\mathcal{P}^{\sigma}_{\zeta} = \rdec^2\left(\frac{H_*}{3\pi \sigma_*}\right)^2.
\eea
{Here}, a star denotes the time when the pivot scale $k_* = 0.05\, \mathrm{Mpc}^{-1}$ crosses the Hubble radius, $H=\dot{a}/a$ is the Hubble scale where a dot denotes derivation with respect to cosmic time, $\epsilon=-\dot{H}/H^2$ is the first slow-roll parameter, $\Mp$ is the reduced Planck mass, and $r_{\mathrm{dec}}$ is a parameter that quantifies the relative energy density of the curvaton at its decay. In the sudden-decay approximation~\cite{Malik:2006pm, Sasaki:2006kq},
\bea 
\rdec \simeq \left( \frac{3\rho_\sigma}{3\rho_\sigma+4\rho_{\mathrm{radiation}}} \right)_{\rm dec}\,.
\eea
This quantity can vary from zero to unity in the case that $\sigma$ dominates the background energy density at the time it decays.

Observations are often discussed in terms of the spectral index $\nS\equiv 1+\dd\ln\calP_\zeta^\mathrm{total}/\dd\ln k$ and the tensor-to-scalar ratio $r\equiv \calP_h/\calP_\zeta^\mathrm{total}$ (where $\calP_h$ is the tensor power spectrum) given by~\cite{Wands:2002bn}
\bea
\label{eq:nsr:slowroll}
\nS-1 & =   \lambda \left(-2\epsilon_*+2\eta_{\sigma*}\right) +\left(1-\lambda\right)\left(-6\epsilon_*+2 \eta_{\phi*}\right)\,, \\
r & = 16\epsilon_* \left(1- \lambda\right)\, .
\eea
Here $\eta_\phi = V_{,\phi \phi}/(3H^2)$ and $\eta_\sigma = V_{,\sigma \sigma}/(3H^2)$, $V_{,x}$ ($V_{,xx}$) denotes the first (second) derivative of $V$ with respect to to $x$, and $\lambda$ denotes the fraction of the total perturbations originating from $\sigma$,
\bea
\label{eq:lambda:def}
\lambda\equiv\frac{{\mathcal P}^{\sigma}_{\zeta}}{{\mathcal P}^{\rm total}_{\zeta}} \, .
\eea
When the primordial density perturbation is entirely due to the spectator field fluctuations then the original curvaton model~\cite{Lyth:2001nq, Enqvist:2001zp, Moroi:2001ct} is realised. Hence, in this work we term situations where $\lambda > 0.9$ as the ``curvaton scenario''. 

At the pivot scale, the latest 2015 BICEP2/Keck Array and Planck \cite{Array:2015xqh, Ade:2015xua} combined observations give ${\mathcal{P}}^{\mathrm{total}}_{\zeta} \sim 2.2\times 10^{-9}$, $\nS =0.9667\pm0.008$ {($95\%$ c.l.)} and $r <0.07$ ($95\%$ c.l.). If the inflaton potential is of the large-field type $U(\phi)\propto\phi^p$, in the curvaton limit $\lambda\simeq 1$, \Eq{eq:nsr:slowroll} implies that $\nS\simeq 1-p/120$, and the observed value of the spectral index means that the inflaton field potential must be close to quartic, $p=4$. The ``simplest'' curvaton scenario with a quadratic inflaton is now disfavoured by the data~\cite{Byrnes:2014xua,Hardwick:2015tma, Vennin:2015egh, Byrnes:2016xlk}. 

The observational constraints on $n_s$ and $r$ imply that when any inflaton potential is included in the analysis, only two classes of models with an additional spectator field are found to be favoured~\cite{Vennin:2015egh}: plateau inflation, which \emph{cannot} fit the data in the curvaton scenario (thereby requiring $\lambda\ll1$), and quartic inflation, which \emph{can only} fit the data in the curvaton scenario ($\lambda\sim1$). An advantage of a quartic potential is that the inflaton field energy decreases like radiation when it oscillates, making the model more predictive by removing the dependence of post-inflationary dynamics on the inflaton decay rate into radiation.% and making the curvaton energy density grow more rapidly.

Another way to detect the curvaton is through primordial non-linearity of the density perturbations, of which the key observable is the local non-Gaussianity of the bispectrum, parametrised by $\fnl$. In the sudden-decay approximation~\cite{Lyth:2005fi, Ichikawa:2008iq}
\bea
  \label{eq:fnl}
\fnl \simeq \lambda^2 \left(\frac{5}{4\rdec}-\frac53-\frac{5\rdec}{6} \right)\,,
\eea
where the observational non-Gaussianity constraint of {$\fnl=0.8\pm5.0$ ($68\%$ c.l.)} \cite{Ade:2015ava} implies that either we predominantly observe inflaton perturbations, $\lambda\simeq0$, or the spectator must have a non-negligible energy density at its decay, $\rdec\gtrsim0.1$.

The contribution from the curvaton to the primordial power spectrum crucially depends on its field value, $\sigma_*$, when observable modes exit the Hubble radius. Combining \Eqs{eq:powerspecs} and~(\ref{eq:lambda:def}), the curvaton dominates the perturbations, $\lambda>1/2$, if $\sigma_*/\Mp < \sqrt{\epsilon_*}\rdec$. Therefore $\sigma_*$ must be sub-Planckian (if it is super-Planckian, it may drive a second phase of inflation and the above formulas do not apply, but we show {in foonote \ref{footnote}} that this case is excluded). The value of $\sigma_*$ is determined by the details of the inflaton's potential $U(\phi)$ over the entire inflating domain, as recently shown in \Ref{Hardwick:2017fjo}. This makes the model more predictive since the typical value of $\sigma_*$ is not a free parameter anymore. This will play an important role in the Bayesian analysis below. In particular, the value of $\sigma_*$ depends on the total duration of inflation, which will allow us to constrain it.\footnote{For a related approach in the context of light vector fields, see \cite{Naruko:2014bxa,Fujita:2017lfu}.}

\section{Priors}

Most previous analyses of curvaton models assumed no knowledge a priori about spectator field values. Instead, we adopt a physical prior for the typical field displacement $\langle \sigma_*^2\rangle^{1/2}$ of the curvaton~\cite{Hardwick:2017fjo}, calculated in the framework of stochastic inflation~\cite{Starobinsky:1986fx}, which describes how, during inflation, the vacuum expectation value (vev) of spectators fields on super-Hubble scales are sourced by quantum fluctuations. This prior depends on the inflaton potential $U(\phi)$ and the total duration of inflation.

In the presence of a plateau inflationary potential (here we choose Higgs inflation~\cite{Bezrukov:2007ep}, whose potential matches the Starobinsky model~\cite{Starobinsky:1980te}, to be the representative member of this class), if inflation lasts more than $N=H^2/m_\sigma^2$ \efolds~\cite{Starobinsky:1986fx, Enqvist:2012xn}, the vev of $\sigma$ reaches a Gaussian equilibrium distribution with vanishing mean, and variance
\bea
\label{eq:hisigmaend}
\left\langle \sigma_*^2 \right\rangle = \frac{3 H_*^4}{8\pi^2 m_\sigma^2}
\,.
\eea
In the presence of a quartic large-field inflationary potential ($U(\phi ) \propto \phi^4$), we find a zero-mean Gaussian with variance
\bea
\label{eq:lfi4sigmaend}
\left\langle \sigma_*^2 \right\rangle =
\left\langle \sigma^2_{\mathrm{in}} \right\rangle +
\frac{H_*^2}{12\pi^2}N_{\mathrm{tot}}^3
\,,
\eea
with a strong dependence upon initial conditions, {as was shown to be the case for a large field inflationary background with monomial power $\ge 2$ in Ref.~\cite{Hardwick:2017fjo}}. {Above}, $N_{\mathrm{tot}}$ is the total number of \efolds~elapsed during quartic inflation and $\langle \sigma^2_{\mathrm{in}} \rangle$ denotes the variance of the curvaton vev distribution at the onset of inflation. 
The distributions (\ref{eq:hisigmaend}) and (\ref{eq:lfi4sigmaend}) define the prior on $\sigma_*$ for plateau and quartic inflation, respectively.

{The expansion history of reheating depends on the mass of the curvaton and the decay rates, $\Gamma$, of the inflaton and the curvaton.\footnote{Here $\Gamma_\phi$ (or $\Gamma_\sigma$) denotes the value of $H$ below which the energy density contained in $\phi$ (or $\sigma$, respectively), or its decay products, redshift like radiation.} Through non-informative priors, (see Appendix) we impose that the onset of the radiation-dominated period occurs after the end of inflation and before the electro-weak symmetry breaking. We also assume that the inflaton and the curvaton decay at least as fast as they would through their minimal coupling to the gravitational sector, $\Gamma > m^3/\Mp^2$~\cite{Enqvist:2013gwf}. As noted earlier, if the inflaton has a quartic potential, its coherent oscillations around the minimum of its potential give rise to a radiation-like era of expansion immediately after inflation~\cite{Turner:1983he}. In this case reheating can be described by two parameters only, the mass and decay rate of the curvaton.}

\section{Results}
\begin{figure}[t]
\begin{center}
\includegraphics[width=0.47\textwidth]{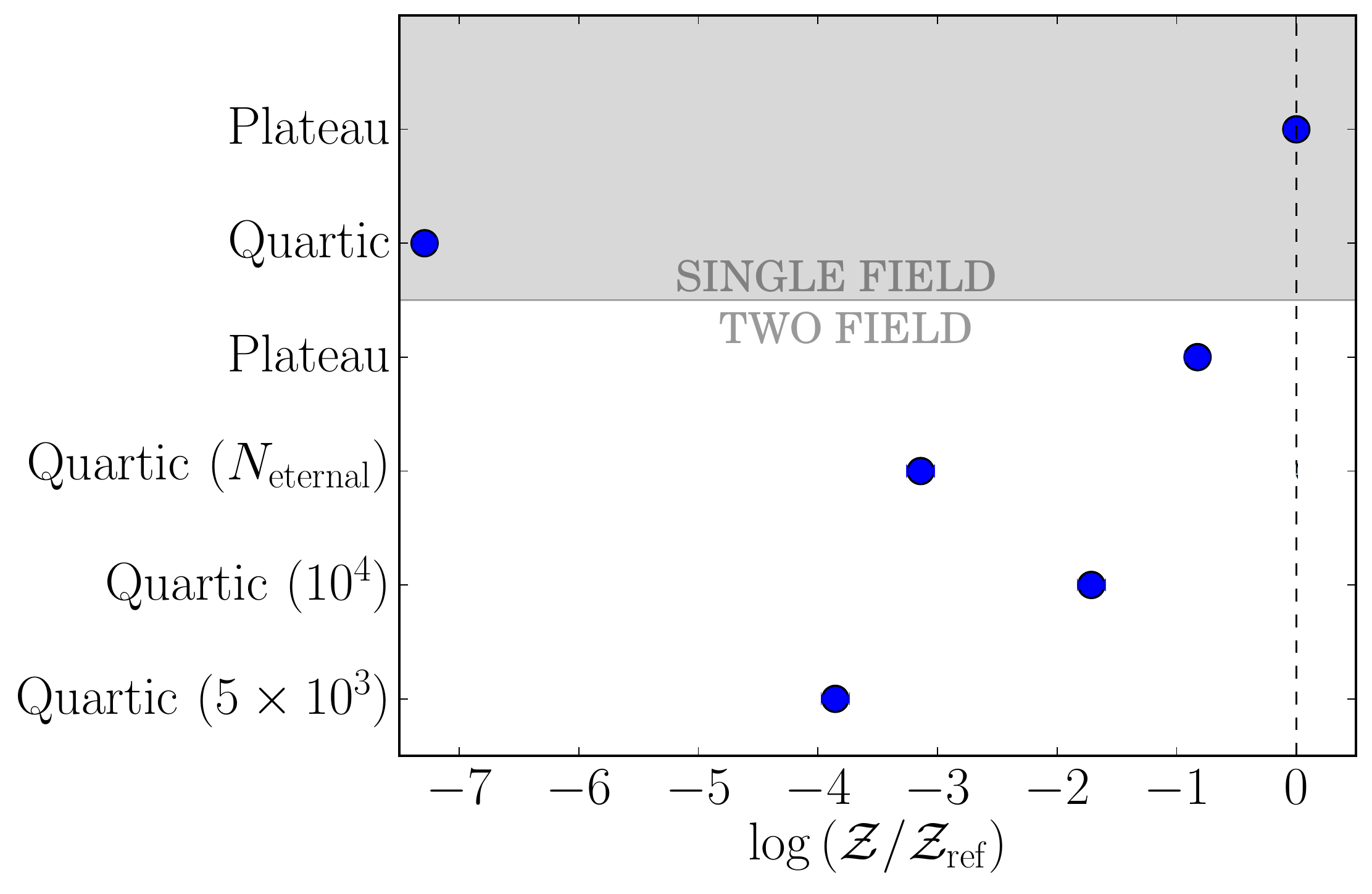}
\end{center}
\caption{Bayesian evidences $\mathcal{Z}$ of the single-field (inflaton) and two-field (inflaton plus spectator) models (inside and below the shaded region, respectively) considered in this work. The plateau model (taken as the reference here) is robust with respect to the introduction of an additional field. Quartic inflation with a spectator field (where the total number of \efolds~is written in parenthesis) has a higher evidence than its single-field version, but lower than the plateau model.}
\label{fig:evidences} 
\end{figure} 

The Bayesian analysis is performed on the January 2015 BICEP2/Keck-Array/Planck data combination \cite{Ade:2015tva}, using the machine-learned effective inflationary likelihood described in \Ref{Ringeval:2013lea}, which has been marginalised over late-time background cosmology, reionisation, and astrophysical foregrounds. The predictions of the models are computed with the curvaton extension of the \texttt{ASPIC} library~\cite{aspic}, making use of the method presented in \Refs{Vennin:2015egh, Hardwick:2016whe}. The Bayesian evidences are integrated using \texttt{MultiNest} \cite{Feroz:2007kg,Feroz:2008xx}; further technical details on the numerical integration can be found in the Appendix. The Bayesian evidences are displayed in \Fig{fig:evidences} and the corresponding posterior distributions in \Fig{fig:observables}. 

\subsection{Single-field versus spectator model}

\begin{figure*}[ht]
  \centering
  \includegraphics[width=0.49 \textwidth]{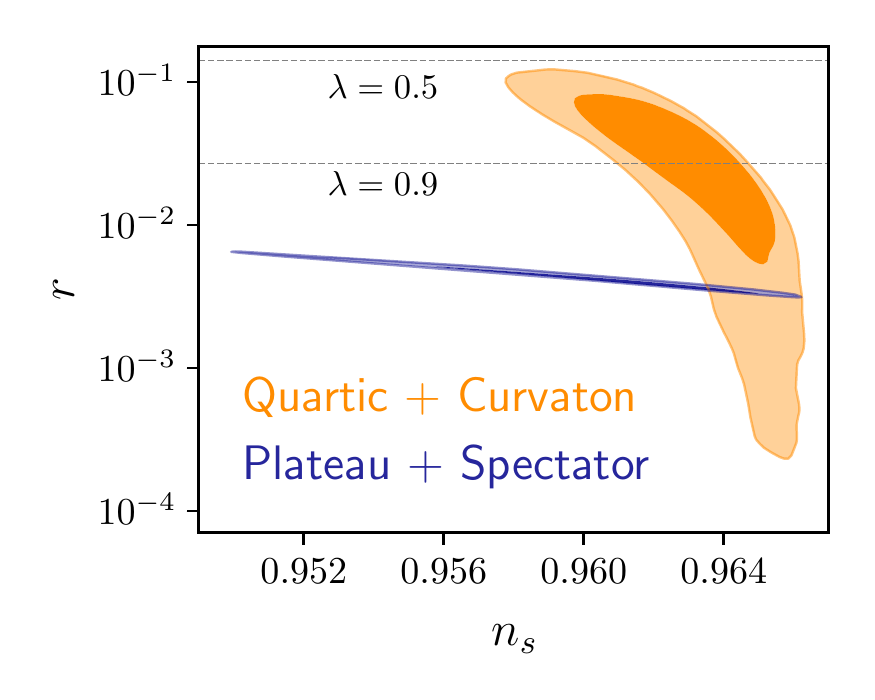}
  \includegraphics[width=0.49 \textwidth]{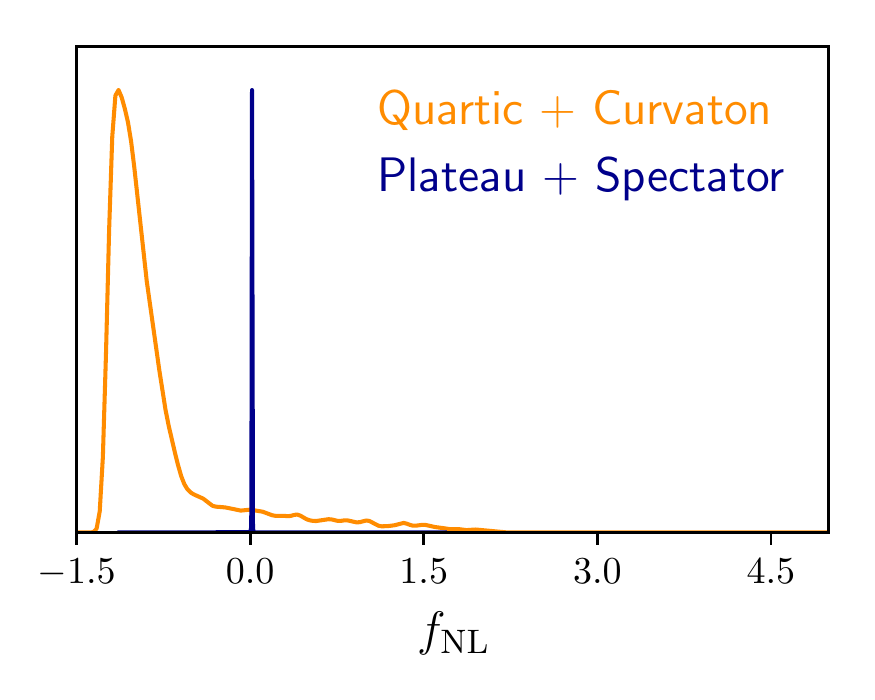} 
  \caption{\label{fig:observables}
Marginal posterior distributions over the key observables from inflation for plateau-like inflation (blue, darker) and quartic inflation (orange, clearer) with a spectator field. In the quartic case, the posterior fraction below the lower (upper) dotted line has more than $90\%$ ($50\%$) of primordial density perturbations generated by the curvaton field. Post-2020 CMB experiments would likely distinguish between or rule out both scenarios in terms of $\nS$ and $r$. In combination with LSS data, the typical value of $\fnl = -5/4$ associated with the curvaton scenario could also be {detected}.}
\end{figure*}

One can check in \Fig{fig:evidences} that for single-field models, plateau potentials are favoured while a quartic potential is strongly disfavoured (and even ruled out at the level of its maximum likelihood). When a light spectator field is included, the evidence of plateau potentials remains stable, and the two-field model cannot be distinguished from its single-field counterpart in terms of its Bayesian evidence~\cite{Jeffreys:1961}. This is because, in spite of the significant enlargement in prior parameter space caused by the introduction of the spectator field, most of the prior mass in the distribution~(\ref{eq:hisigmaend}) reproduces single-field phenomenology, which gives a very good fit to the data irrespective of the value of the reheating parameters. This result is consistent with \Refs{Vennin:2015egh, Hardwick:2016whe}.

%-----------------

%For single-field models, plateau potentials are favoured while a quartic potential is strongly disfavoured (even ruled out at the level of its maximum likelihood), see \Fig{fig:evidences}. The evidence for plateau potentials remains stable when a light spectator field is included, so its presence cannot be discerned in terms of Bayesian evidence~\cite{Jeffreys:1961}. This is because, in spite of the significant enlargement in prior parameter space caused by the introduction of the spectator field, most of the prior mass in the distribution~(\ref{eq:hisigmaend}) reproduces single-field phenomenology, which gives a very good fit to the data irrespective of the value of the reheating parameters. This result is consistent with what was found in \Refs{Vennin:2015egh, Hardwick:2016whe}.

%-------------------

For the quartic potential, the evidence obtained once a spectator field is included depends on the total duration of inflation, $N_{\mathrm{tot}}$, {and the variance of the curvaton vev distribution at the onset of inflation, $\langle \sigma^2_{\mathrm{in}} \rangle$}, through the prior distribution~(\ref{eq:lfi4sigmaend}) for the curvaton vev. We give the Bayesian evidence for a few values of $N_{\mathrm{tot}}$ in \Fig{fig:evidences}, {taking $\langle \sigma^2_{\mathrm{in}} \rangle=0$, and} $N_{\mathrm{tot}}\sim 6\times 10^4$ as an upper bound, since for larger values the inflaton would initially be in the ``self-reproducing'' regime~\cite{Linde:2005ht, Winitzki:2008zz} where stochastic corrections to its dynamics become important and the calculation of \Ref{Hardwick:2017fjo} does not apply; {however, we are protected from this limit by the fact that it is disfavoured observationally, since it would locate most of the prior mass in spectator vevs so large that they drive a second phase of quadratic inflation}.\footnote{If the light spectator field is displaced by $\sigma_*^2\gtrsim 2\Mp^2$ during inflation, then it may drive a second period of inflation, which lasts for $N_2\simeq \sigma_*^2/(4 \Mp^2)$ \efolds. The amplitude of the curvaton perturbations generated during the first period of inflation is~\cite{Vernizzi:2006ve}
\bea
{\mathcal{P}}^{\sigma}_{\zeta} = N_2\left(\frac{H_*}{2\pi\Mp}\right)^2\, .
\eea
Independently of the inflaton potential, the tensor-to-scalar ratio is given by
\bea 
r=\frac{{\mathcal{P}}_h}{{\mathcal P}^{\sigma}_{\zeta}+{\mathcal P}^{\phi}_{\zeta}} = \lambda \frac{{\mathcal P}_h}{{\mathcal P}^{\sigma}_{\zeta}} = \lambda \frac{8}{N_2}\, , 
\eea
where ${\mathcal P}_h=8 [H_*/(2\pi \Mp)]^2$. The observational bound on $r$ then imposes
\bea 
\lambda \lesssim \frac12 \left(\frac{N_2}{60}\right)\, .
\eea
Since we require $N_2<60$, because otherwise the first period of inflation would end before the observable modes exit the horizon, this implies that a quadratic spectator field that then inflates the Universe cannot generate the majority of the observed perturbations.\label{footnote}}

In all cases, quartic models with a spectator field are favoured with respect to their single-field counterpart, but are still moderately or strongly disfavoured with respect to the plateau potential.

In terms of the observables shown in \Fig{fig:observables}, plateau inflation (the Higgs inflation or Starobinsky model in the present case) with a spectator field presents similar phenomenology to its single-field counterpart, namely a small tensor-to-scalar ratio and a slow-roll suppressed value for $\fnl$ that is currently (and in the foreseeable future) undetectable. For quartic inflation, independently of the duration of inflation, the tensor-to-scalar ratio and the spectral index are correlated, with bluer spectra corresponding to reduced gravitational waves, and non-Gaussianity has the typical amplitude $\fnl\simeq-5/4$, which, from \Eq{eq:fnl}, corresponds to a preference for values $\lambda\simeq \rdec \simeq 1$, \ie to situations where the curvaton dominates the energy budget of the Universe when it decays and provides the dominant contribution to primordial density perturbations.

Post-2020 CMB experiments \cite{Matsumura:2013aja, Finelli:2016cyd, DiValentino:2016foa} will shrink the $1$-sigma constraints on the inflationary observables to $\Delta \nS \sim 2 \times10^{-3}$ and $\Delta r\sim 10^{-4}$, while cross-correlation with future LSS experiments should drive the constraint on local non-Gaussianity down to $\Delta \fnl\sim 0.4$ \cite{Schmittfull:2017ffw}. This would be enough to distinguish between plateau inflation (with or without a spectator field) and quartic inflation with a curvaton, or even to rule out both models.

\subsection{Measuring the {maximum} duration of inflation}

For quartic potentials with a spectator field, the data shows strong preference for curvatonic phenomenology, which corresponds to sub-Planckian spectator field values of a few $10^{-2}\,M_\mathrm{Pl}$. This yields an ``optimal'' value for {the variance of the prior distribution~(\ref{eq:lfi4sigmaend})}, such that it maximises the parameter volume that falls within this range of values.

A smaller variance limits the spectator field vev so that single-field quartic inflation is recovered, which is ruled out observationally. A larger variance locates most of the prior mass in spectator vevs so large that they drive a second phase of quadratic inflation, which is also ruled out {(see footnote \ref{footnote}).}

{Let us assume $\langle \sigma^2_{\mathrm{in}} \rangle=0$ in Eq.~\eqref{eq:lfi4sigmaend}. In that case,} using Bayes theorem, the posterior on the total duration of inflation can be computed according to
\bea
\label{eq:nmaxpost}
\mathcal{P}\left(N_\mathrm{tot}\right\vert\left.\mathcal{D}\right) \propto
\mathcal{P}\left(\mathcal{D}\right\vert\left.N_\mathrm{tot}\right)\,\pi\left(N_\mathrm{tot}\right)
\, ,
\eea
where $\mathcal{P}(\mathcal{D}|N_\mathrm{tot}) = \mathcal{Z}(N_\mathrm{tot})$ is the evidence of the quartic plus spectator field model with prior~(\ref{eq:lfi4sigmaend}) on $\sigma_*$ corresponding to $N_\mathrm{tot}$, and $\pi(N_\mathrm{tot})$ is the prior we set on the duration of inflation. 

\begin{figure}[t]
  \centering
  \includegraphics[width=0.48\textwidth]{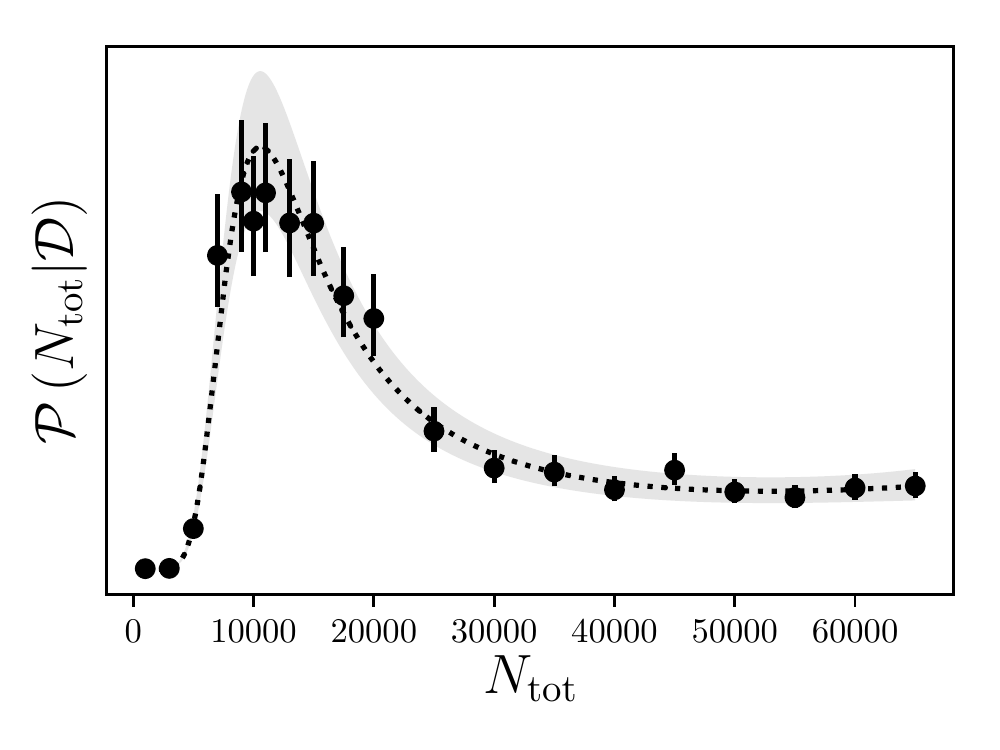}
  \caption{\label{fig:Nefolds}
    Marginal posterior over the total number of \efolds~of inflation $N_{\mathrm{tot}}$ with a uniform prior, for quartic inflation with a spectator field, {assuming negligible initial spectator field values (otherwise this plot represents an upper bound on $N_\mathrm{tot}$)}. The upper limit corresponds approximately to the ``self-reproducing'' regime, $N_\mathrm{tot}\sim 6\times 10^4$. The dotted line and the grey band are respectively the mean and $1$-sigma confidence-level limit of a logarithmic Gaussian process interpolation with maximum-a-posteriori noise level, scale and correlation length~\cite{Rasmussen:2005:GPM:1162254}. The black dots and bars are the evidences and their error computed with \texttt{MultiNest}.}
\end{figure}

We reconstruct this posterior in \Fig{fig:Nefolds}, where one can see that inflation is constrained to last less than a few tens of thousands of \efolds. In particular, cases where inflation starts close to the ``self-reproducing'' regime ($N_{\mathrm{tot}}\sim 6\times 10^4$), are strongly disfavoured~\cite{Winitzki:2010yz}. This is because in such cases, \Eq{eq:lfi4sigmaend} yields $\langle \sigma_*^2\rangle^{1/2}> \Mp$ (which is true in any large-field inflationary potential~\cite{Hardwick:2017fjo}) and the spectator field drives a second phase of inflation.

If  the initial variance $\langle\sigma^2_{\rm in}\rangle$ does not vanish then the constraint that we have obtained {becomes} an upper bound on the duration of inflation, {and the conclusion that inflation should not start in the self-reproducing regime becomes even stronger.}

\section{Conclusion} \label{sec:con}

If inflation is realised in the presence of light spectator fields, which appear in many high-energy embeddings of inflation, then those fields may source part or all of the primordial density perturbations. Recent CMB measurements have now reached a level of accuracy such that there is no inflationary potential for which the single-field limit, where all perturbations come from the inflaton field, and the curvaton limit, where all perturbations come from a spectator field, are both allowed by the data. This is why, for the first time, we are in a position where a Bayesian model comparison of inflationary models with spectator fields yields non-trivial results.

For instance, we have found that if the potential is of the plateau type, the single-field limit is the preferred one (the predictions of the model are robust under the introduction of a spectator field), while quartic potentials are favoured only in the curvaton limit. Both {combinations} are equally favoured by current data, but we have shown that future CMB and LSS measurements will allow us to distinguish between them.

The contribution from spectator fields to cosmological perturbations strongly depends on their field values at the end of inflation \cite{Hardwick:2017fjo}. The accumulation of long-wavelength quantum fluctuations during the entire inflationary period gives rise to a distribution for the local field displacement that depends on the total duration of inflation {$N_{\mathrm{tot}}$, plus a possible contribution at the onset of inflation $\langle \sigma^2_{\mathrm{in}} \rangle$. As a consequence, a combination of both parameters can be constrained by the data.}

In the curvaton limit, the inflationary potential is constrained to be close to the quartic type. In that case, {spectator field values that are too small would fail to source cosmological perturbations, and too large values would drive} a second phase of inflation, {which is observationally disfavoured}.

{Assuming $\langle \sigma^2_{\mathrm{in}} \rangle=0$}, we compute the posterior distribution on $N_{\mathrm{tot}}$ (see \Fig{fig:Nefolds}), and find that inflation could not last more than a few tens of thousands of \efolds. In particular, it is very unlikely that one starts quartic inflation in the so-called ``self-reproducing'' regime. {Letting $\langle \sigma^2_{\mathrm{in}} \rangle>0$ makes that upper bound even stronger.}

For the first time, we have thus quantified how much cosmological data can constrain the pre-inflationary history, much beyond the $N \gtrsim 60$ epoch probed by potential large scale CMB anomalies. One should note that the mechanism we presented is not only sensitive to the duration of inflation but also on the shape of the inflationary potential over its entire inflating domain, and on the spectator field displacement prior to inflation. This opens up a new observational window that extends the conventional scales by orders of magnitude and allows us to explore the physics of the very early Universe beyond our currently observable horizon.

\smallskip\smallskip\smallskip
\acknowledgments{{\bf Acknowledgements} CB is funded by a Royal Society University Research Fellowship. JT acknowledges funding from the European Research Council under the European Union's Seventh Framework Programme (FP/2007-2013) / ERC Grant Agreement No.~[616170]. VV acknowledges funding from the European Union's Horizon 2020 research and innovation programme under the Marie Sk\l odowska-Curie grant agreement N${}^0$ 750491. VV and DW acknowledge financial support from UK Science and Technology Facilities Council grant ST/N000668/1. RH is supported by UK Science and Technology Facilities Council grant ST/N5044245.}

\appendix

\section{Appendices}

\subsection{Details on the priors and the computation of the evidences}\label{app:stats}

The non-informative prior described in the paper on the reheating parameters, the mass of the curvaton and the decay rates $\Gamma$ of the inflaton and the curvaton, is 
\bea
\label{eq:newbaseline}
\Gamma_\sigma &\sim \log \uniform\left[\max\left(H_\mathrm{EW}, \frac{m_\sigma^3}{\Mp^2}\right), \min\left(H_\uend, m_\sigma\right)\right]\, ,\\
\Gamma_\phi   &\sim \log \uniform\left[\max\left(H_\mathrm{EW}, \frac{H_\mathrm{end}^3}{\Mp^2}\right), H_\uend\right]\,, \\
m_\sigma &\sim \log \uniform\left[ H_{\mathrm{EW}}, H_\uend \right]\, ,
\eea
where $H_\uend$ is the Hubble scale at the end of inflation, $H_\mathrm{EW}=(150\,\mathrm{GeV})^2/\Mp$ is the Hubble scale at electro-weak symmetry breaking, and $x\sim\log \mathcal{U}[a,b]$ means that $\log x$ is uniformly distributed between $\log a$ and $\log b$.

In the quartic inflaton case, the radiation-like reheating of the inflaton, described as $\Gamma_\phi=H_\uend$, is imposed via a half log-normal $\log_{10}(\Gamma_\phi) \sim \normal_{1/2}\left[\log_{10}(H_\mathrm{end}),(1/2)^2\right]$. This needs to be done for numerical purposes, since $H_\mathrm{end}$ is a derived quantity that depends of the full parameter combination and can only be computed a posteriori.

In the models presented in this paper, the total power of the primordial density perturbations constitutes an additional free parameter, which we have omitted because it affects both models equally. For numerical purposes, we use a log-uniform prior which comfortably contains the posterior observed by Planck for this parameter. Thus, the total parameter space sampled is $(\Gamma_\phi, \Gamma_\sigma, m_\sigma, \sigma_\mathrm{end}, A_\mathrm{s})$, and our posteriors and evidences are conditioned to the model producing \emph{close to} the right amount of power.

We ensure the correct normalisation of the evidences by dividing the marginal likelihood by the total prior mass in the same parameter domain, obtained with a quick \texttt{MultiNest} integration of a mock unit likelihood. All results are obtained with 1000 \emph{live points} and a very low \emph{sampling efficiency} of 0.01 (i.e.\ inverse of ellipsoid enlargement factor). A significant enlargement of the ellipsoids is needed to properly account for the hard edges of the prior and the fact that in the quartic case the mode of the spectator field value is located at the edge of the prior (otherwise if a mode at the edge of the prior is partially or totally missed by the initial sample of live points, the final evidence will be undervalued). This low efficiency produces a lot of rejected points that spoil the computation of the weights used by the Importance Nested Sampling estimator \cite{2013arXiv1306.2144F}, what makes it numerically unstable, most often severely undervalued. Thus, we use the standard nested sampled estimator in this paper.

\bibliography{single-vs-spectator}

\end{document}